\documentclass[aps,showpacs,preprintnumbers,prd,twocolumn,nofootinbib,superscriptaddress]{revtex4}

\IfFileExists{srcltx.sty}{\usepackage[active]{srcltx}}

\usepackage{epsfig}
\usepackage{graphicx}
\usepackage{epsfig}
\usepackage{bm}
\usepackage{amsmath}
\usepackage{amssymb}
\usepackage{multirow}

\newcommand{\beq}{\begin{equation}}
\newcommand{\eeq}{\end{equation}}
\newcommand{\bea}{\begin{eqnarray}}
\newcommand{\eea}{\end{eqnarray}}

\def\MeV{\: {\rm MeV}}
\def\GeV{\: {\rm GeV}}

\newcommand{\slashed}[1]{{#1}\hspace{-2mm}/}

\def\simle{\lower 2pt \hbox {$\buildrel < \over {\scriptstyle \sim }$}}
\def\simge{\lower 2pt \hbox {$\buildrel > \over {\scriptstyle \sim }$}}

\begin{document}

\preprint{UCLA/08/TEP/18}

\title{Heavy sterile neutrinos and supernova explosions}

\author{George M. Fuller}
\affiliation{Department of Physics, University of California, San Diego, La Jolla, CA 92093-0319}

\author{Alexander Kusenko}
\author{Kalliopi Petraki}

\affiliation{Department of Physics and Astronomy, University of California, Los Angeles, CA
90095-1547, USA
}


\begin{abstract}
We consider sterile neutrinos with rest masses $\sim 0.2\,{\rm GeV}$ and with vacuum flavor mixing angles $\sin^2\theta >10^{-8}$ for mixing with $\tau$-neutrinos, or $10^{-8}<\theta^2 <10^{-7}$ for mixing with muon neutrinos.  Such sterile neutrinos could augment core collapse supernova shock
energies by enhancing energy transport from the core to the vicinity of the shock front. The decay of these neutrinos could produce a flux of very energetic active neutrinos, detectable by future neutrino
observations from galactic supernova. The relevant range of sterile neutrino masses and mixing
angles can be probed in future laboratory experiments.

\end{abstract}

\pacs{14.60.St, 95.35.+d}

\maketitle

\section{Introduction}

Neutrino masses are usually incorporated into the Standard Model (SM) by the addition of
SU(3)$\times$SU(2)$\times$U(1) singlet fermions, often called right-handed neutrinos~\cite{seesaw}.
These gauge singlets can have Majorana masses  as low as
a few eV~\cite{deGouvea:2005er}, or as large as the Grand Unified scale~\cite{seesaw}. If the
Majorana mass terms are large, the particles associated with the singlet fields are very heavy.
However, if the Majorana masses are below the electroweak scale, the corresponding degrees of
freedom appear in the low-energy effective theory as so called \textit{sterile neutrinos}.
Light sterile neutrinos with masses of a few keV could be the cosmological dark
matter~\cite{dw,nuMSM}, their production in a supernova could result in large pulsar
kicks~\cite{pulsars}, and they could affect supernovae in a variety of  ways~\cite{supernova_misc}.
The same particles can play an important role in the formation of the first stars~\cite{reion},
baryogenesis~\cite{baryogenesis}, and other astrophysical phenomena~\cite{Biermann:2007ap}.

In this paper we investigate the effect of heavier sterile neutrinos in astrophysical systems. We
show that sterile neutrinos with masses $\sim 0.2 \GeV$ and small mixing $\sin^2 \theta \sim
10^{-8}$, with either the muon or tau neutrinos, could be produced in supernova cores and
subsequently augment, via their decay, the energy transport from the core to the region around the
stalled shock, thereby increasing the prospects for a core collapse supernova explosion. This
scenario could be testable, both by future laboratory searches for heavy sterile neutrinos and by
observations of the neutrino signal from supernovae.

Our analysis differs from earlier work. Heavy neutral leptons, with masses 10~keV -- 10~MeV,
produced in supernovae, have been considered for setting limits and for powering supernova
explosions~\cite{Lepton Decays in SN}. 
However, the mass range we consider here, 145 -- 250 MeV, is qualitatively different in that the
sterile neutrinos decay predominantly into a pion and a light fermion. If the mixing angle with the
electron neutrino is negligible, and the sterile neutrino mass $M_s$ is in the range $m_{\pi^0} <
M_s < (m_{\pi^0}+m_\mu)$, the daughter pion is the neutral pion, which decays into  two photons: 
$\nu^{\rm (s)} \rightarrow \nu^{\rm (a)} \pi^0 \rightarrow \nu^{\rm (a)} \gamma
\gamma$~\cite{Dolgov:2000jw}, where $a=(\mu,\tau) $.  This decay mode, with lifetime $\sim 0.1\,{\rm
s}$, changes the impact of sterile neutrinos on the supernova  explosion. To distinguish sterile
neutrinos that decay mainly into photons from the other types, we call them {\em
eosphoric}.\footnote{From the ancient Greek god $E \omega\sigma\varphi \acute{o} \rho o \varsigma$,
the bearer of light.}
As a consequence of small mixing angles, $\sin^2 \theta \sim 10^{-8} - 10^{-7}$, the eosphoric sterile neutrino production and propagation history inside the supernova environment could be significantly different from those of the neutral leptons with electroweak scale interactions which have been considered previously \cite{Lepton Decays in SN}.

We assume that the particle physics lagrangian at low energies comprises the Standard Model, albeit modified as indicated. The Standard Model was originally formulated with massless neutrinos $\nu_i$ transforming as components of the electroweak SU(2) doublets $L_\alpha$ ($\alpha =1,2,3$), but here we will extend it to include seesaw mass terms for neutrinos~\cite{seesaw}, in which we allow the Majorana masses to be below the electroweak scale:
\beq  {\cal L} = {\cal L_{\rm SM}} + i \bar
N_a \slashed{\partial} N_a - y_{\alpha a} H^{\dag} \,  \bar L_\alpha
N_a - \frac{M_a}{2} \; \bar N_a^c N_a + h.c.
\label{L}
\eeq
The neutrino mass eigenstates
$\{ \nu^{\rm (a)}_1 \nu^{\rm (a)}_2 \nu^{\rm (a)}_3, \nu^{\rm (s)}_1, \nu^{\rm (s)}_2, ..., \nu^{\rm (s)}_n \}$  are linear combinations of the weak eigenstates $\{\nu_\alpha, N_a \}$.  The states $ \nu^{\rm (a)}_{1,2,3}$ are {\em active} and have masses below 0.2~eV, while $ \nu^{\rm (s)}_{1,...,n}$ are {\em sterile}. In particular, several recent studies focus on the $\nu$MSM~\cite{nuMSM}, a model with $n=3$ sterile neutrinos: one with  a few ${\rm keV}$ rest mass (dark matter), and two nearly degenerate, heavier states. This model facilitates leptogenesis via neutrino oscillations~\cite{baryogenesis}.

We will consider sterile neutrinos with masses 145--250 MeV, and with vacuum flavor mixing angle $\sin^2\theta \sim 10^{-8} - 10^{-7}$ for mixing with either muon or tau neutrinos in vacuum. The current bounds~\cite{sterile_constraints} allow this range of mixing angles with both muon and tau neutrinos.  For sterile neutrinos mixing only with tau neutrinos an even broader range is allowed~\cite{sterile_constraints,sterile_exp_constraints}. Sterile neutrinos with somewhat smaller vacuum mixing could produce a pre-nucleosynthesis matter-dominated epoch in the early universe, and this is being investigated separately~\cite{us-BBN}. In section \ref{production}, we investigate the production of these neutrinos in the core of a hot proto-neutron star. In section \ref{shock}, we discuss how the decay of eosphoric neutrinos leads to energy deposition in the mantle above the post-collapse neutron star which could help in shock revival. Section \ref{nu signal} is devoted to the neutrino signal produced by these decays, and the associated observational signature of this scenario. We show that this is consistent with the neutrino detection from SN1987A.

\section{Sterile Neutrino Production \label{production}}

Sterile neutrinos are produced in a hot proto-neutron star by electron-positron or neutrino pair annihilation, and by inelastic scattering of $\nu_{\mu (\tau)}$ and $\bar{\nu}_{\mu (\tau)}$ on any of the fermionic species. Since $n,p,e^-,e^+,\nu_e,\bar{\nu}_e$ are degenerate in the core, fermi-blocking will render pair annihilation of and inelastic scattering on the non-degenerate neutrino species $\nu_\mu, \nu_\tau$  the dominant process for $N_s$ production~\cite{production_processes}. The amplitude for these processes, in terms of the Mandelstam variables, is~\cite{Hannestad:1995rs}:

\begin{table}[h!]
\centering
\begin{tabular}{c|c}
\hline \hline
Process &  Amplitude  \\
\hline
 $\bar{\nu}_{\mu (\tau)}  +  \nu_{\mu (\tau)}   \rightarrow \bar{\nu}_{\mu (\tau)}  + \nu_s $   &  $\sin^2 \theta \  32 G_F^2 \ u (u-M_s^2)$ \\
 $\bar{\nu}_{\tau (\mu)}  +  \nu_{\tau (\mu)}   \rightarrow \bar{\nu}_{\mu (\tau)}  + \nu_s $   &  $\sin^2 \theta \ \ 8 G_F^2 \ u (u-M_s^2)$ \\
 $\nu_{\tau (\mu)}        +  \nu_{\mu (\tau)}   \rightarrow       \nu_{\tau (\mu)}  + \nu_s $   &  $\sin^2 \theta \ \ 8 G_F^2 \ s (s-M_s^2)$ \\
 $\bar{\nu}_{\tau (\mu)}  +  \nu_{\mu (\tau)}   \rightarrow \bar{\nu}_{\tau (\mu)}  + \nu_s $   &  $\sin^2 \theta \ \ 8 G_F^2 \ u (u-M_s^2)$ \\
\hline \hline
\end{tabular}
\caption{Dominant processes contributing to sterile neutrino production in the supernova core, for mixing with the muon (tau) neutrino by angle $\theta$. Since the sterile neutrino is its own antiparticle, the charge-conjugated counterparts of the above processes also contribute to $\nu_s$ production.}
\label{processes}
\end{table}

We have calculated the sterile neutrino production rate numerically, by appropriate phase-space integrations involving particle distributions and blocking factors~\cite{Hannestad:1995rs}. Sterile neutrinos with masses $M_s \approx 145-250$~MeV and vacuum mixing $\sin^2\theta \sim 10^{-8} - 10^{-7}$ are not trapped; they stream freely out of the supernova core. Since both the production cross sections $\sigma \sim \sin^2 \theta \, G_F^2 \, E^2 \propto T^2$ and the $\nu_{\mu (\tau)}$ number density $n\sim T^3$ grow rapidly with temperature, the sterile neutrino flux and luminosity depend strongly on temperature. Our numerical results, for a sterile neutrino with mass $M_s = 200 \MeV$, and for temperatures in the range 15-70 MeV, are well approximated by the following fitting functions:
\beq
\frac{d N_s}{dt \ d V} \approx 7 \times 10^{52} {\rm \frac{1}{s \ km^3}} \left(\frac{\sin^2 \theta}{5 \times 10^{-8}} \right)
\left(\frac{T}{35 \, \MeV}\right)^{6.8} e^{-\frac{M_s}{T}}
\label{flux}
\eeq
and
\beq
\frac{d L_s}{d V} \approx 3 \times 10^{49} {\rm \frac{erg}{s \ km^3}} \left(\frac{\sin^2 \theta}{5 \times 10^{-8}} \right)
\left(\frac{T}{35 \; \MeV}\right)^{7.2} e^{-\frac{M_s}{T}}
\label{luminosity}
\eeq

Supernova cores typically reach peak temperatures $T \sim 35 \MeV$, at $R \sim 10-20$~km, for a time interval of a few seconds~\cite{Prakash:1996xs}.
Equations (\ref{flux}) and (\ref{luminosity}) then imply that the sterile neutrino burst will last 1-5 s and drain total energy $E_s \sim 10^{51}$~erg. Sterile neutrinos will be emitted with average Lorentz factor
\beq
\gamma_s \sim \left. \frac{d L_s}{d V} \right/ M_s \frac{d N_s}{dt \ d V} \approx 1.5
\label{ gamma}
\eeq
which is rather insensitive to temperature within the relevant range.

\begin{figure}
  \centering
  \includegraphics[width=7.5cm]{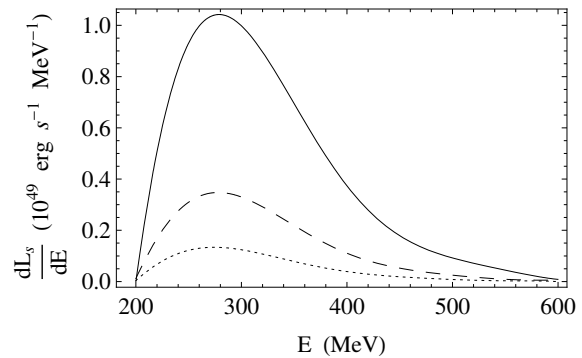}
  \caption{Energy spectrum of sterile neutrinos, of mass $M_s = 200 \MeV$ and mixing angle $\sin^2
\theta = 5 \times 10^{-8}$, produced in the temperature profiles shown in Fig.9 of
Ref.~\cite{Pons:1998mm}.}
  \label{Rates_Pons}
\end{figure}

\begin{figure}
  \centering
  \includegraphics[width=7.5cm]{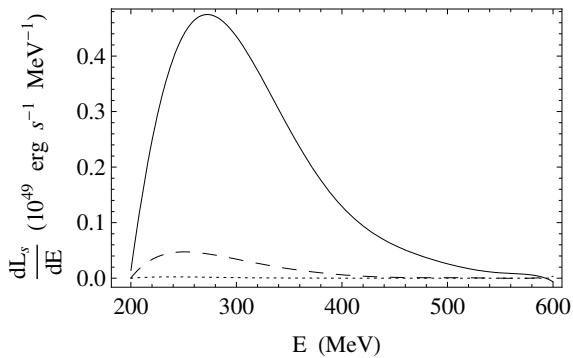}
  \caption{Energy spectrum of sterile neutrinos, of mass $M_s = 200 \MeV$ and mixing angle $\sin^2
\theta = 5 \times 10^{-8}$, produced in model ${\rm  s15Gio\_1d.b}$  of Ref.~\cite{Buras:2005rp}
(shown in Fig.20 of  Ref.~\cite{Buras:2005rp})}
  \label{Rates_Buras}
\end{figure}

In particular, we employ models of supernova cores from 1-dimensional simulations: the model GM3np of a proto-neutron star, from Ref.~\cite{Pons:1998mm}, and the model s15Gio$\_$1d.b, from Ref.~\cite{Buras:2005rp}, for a star with main-sequence mass $15 \ M_\odot$. This simulation implements general-relativistic gravity, advanced description of neutrino interactions and full spectral treatment of neutrino transport, but it fails to generate a successful explosion.
We reproduce the temperature profiles of these models 
and present the resulting sterile neutrino luminosity-energy spectra in figures \ref{Rates_Pons} and
\ref{Rates_Buras}. Tables \ref{Lumin_Pons}, \ref{Lumin_Buras} contain our numerical results for the
total luminosity in heavy sterile neutrinos.

\begin{table}
\centering
\begin{tabular}{c|c|c|c}
\hline \hline
$M_s$          &  \multicolumn{3}{|c}{$L_s$ (erg/s)} \\
\cline{2-4}
(MeV)          &  1s                     &  5s                   & 10s \\
\hline
150            &  $ 5  \times 10^{51}$ & $ 2 \times 10^{51}$ & $ 6 \times 10^{50}$ \\
200            &  $ 2  \times 10^{51}$ & $ 8 \times 10^{50}$ & $ 3 \times 10^{50}$ \\
250            &  $ 1  \times 10^{51}$ & $ 4 \times 10^{50}$ & $ 1 \times 10^{50}$ \\
\hline \hline
\end{tabular}
\caption{Total luminosity in sterile neutrinos, for the temperature profiles of model GM3 of
Ref.~\cite{Pons:1998mm}. The mixing angle is taken to be $\sin^2 \theta = 5 \times
10^{-8}$.}
\label{Lumin_Pons}
\end{table}

\begin{table}
\centering
\begin{tabular}{c|c|c|c}
\hline \hline
$M_s$          &  \multicolumn{3}{|c}{$L_s$ (erg/s)} \\
\cline{2-4}
(MeV)          &  72.5ms             &  113.7ms             & 170.7ms \\
\hline
150            &  $2 \times 10^{49}$ &  $3 \times 10^{50}$  & $ 2 \times 10^{51}$ \\
200            &  $4 \times 10^{48}$ &  $1 \times 10^{50}$  & $ 1 \times 10^{51}$ \\
250            &  $6 \times 10^{47}$ &  $3 \times 10^{49}$  & $ 4 \times 10^{50}$ \\
\hline \hline
\end{tabular}
\caption{Total luminosity in sterile neutrinos, for the temperature profiles of ${\rm 
s15Gio\_1d.b}$  of Ref.~\cite{Buras:2005rp}. The mixing angle is taken to be
$\sin^2 \theta = 5 \times 10^{-8}$.}
\label{Lumin_Buras}
\end{table}

These calculations show that sterile neutrinos in the parameter range under consideration can play an important role in the energy transport and deposition budget in the supernova environment. Although eosphoric sterile neutrino production is suppressed by their small mixing angles and relatively large rest masses, their potential impact is enhanced by their free-streaming out from the core. The net result, for typical supernova cores, is a total energy output in sterile neutrinos of the order of the initial bounce-shock energy, $\sim 10^{51}$~erg. As a result, eosphoric sterile neutrinos carry an amount of energy that, if deposited under or in the vicinity of the shock, could have an impact on the supernova explosion mechanism. 

\section{Supernova Shock Enhancement \label{shock}}

In standard gravitational collapse, a shock forms at the edge of a cold ($T \approx 2\,{\rm MeV}$), low entropy, homologous \lq\lq inner\rq\rq\ core piston and then subsequently propagates outward through the outer core, heating this region, but losing energy through  photo-dissociation of heavy nuclei into free nucleons and alpha particles and, as a consequence, in $\sim 100\,{\rm ms}$ evolves into a \lq\lq stalled\rq\rq\ standing accretion shock. With standard weak interaction physics, this shock may be revived by neutrino heating or neutrino heating aided by multi-dimensional hydrodynamic effects, ultimately producing a supernova explosion~\cite{Bethe:1990mw,Buras:2005rp}. The inner core heats up and the whole core de-leptonizes on a time scale $\sim 1\,{\rm s}$, the neutrino diffusion time scale.

The eosphoric neutrinos  decay into $\gamma$-ray photons and very energetic ($\sim 80\,{\rm MeV}$)
active neutrinos outside the neutron star but inside the envelope.  These decays increase the total energy
of the envelope, making it easier to eject the material from the gravitational potential well of the
neutron star. Their decay time is~\cite{Dolgov:2000jw} 
\beq
\tau_s = \left[\frac{G_F^2 M_s (M_s^2-m_\pi^2) f_\pi^2 \sin^2 \theta}{16 \pi}  \right]^{-1} 
\label{Decay Rate}
\eeq
where $f_\pi = 131 \MeV$. 
For mixing with $\nu_\mu$, in the window $M_s = 145 - 250 \MeV$ and $\sin^2 \theta = 10^{-8} - 10^{-7}$, their decay time is $\tau_s \sim 0.1 \ s$.
Emitted from the core with $\gamma_s \sim 1.5$, they would deposit their energy at distances $R_s \sim 10^9$~cm.
For mixing with $\nu_\tau$, a much wider parameter range is allowed. A requirement that the energy output in sterile neutrinos remains $E_s \sim 10^{51}-10^{52}$~erg, essentially to stay within observational supernova bounds, would limit the parameter region of interest, but it would still allow for lifetimes as small as $\tau_s \sim 10^{-3}$~s and $R_s \sim 10^7$~cm. From eqs. (\ref{luminosity}) and (\ref{Decay Rate}), we see that this can be the case, for example, if $M_s \sim 300 \MeV$ and $\sin^2 \theta \sim 10^{-6}$.

The $\gamma$-ray photons quickly thermalize with matter and could generate a pressure gradient
of sufficient magnitude to affect the standard picture of the gravitational collapse. The energy density deposited in photons is $u_\gamma =
(L_s/4\pi R^2 \beta_s) (1 - e^{-R/R_s})$ and the resulting pressure gradient is
\beq
- \frac{dP_\gamma}{dR} \sim - \frac{d u_\gamma}{dR}
= \frac{L_s}{2 \pi \beta_s R^3} \left[1 - \left(1 + \frac{R}{2 R_s} \right) e^{-R/R_s} \right].
\label{pressure}
\eeq
%

An amount of energy $E_s \sim 10^{51}-10^{52}$~erg, deposited in a region with $M \sim 0.1\,{\rm M}_\odot$ could give an increase in entropy-per-baryon $\Delta s\sim E_s/(T M/m_p)\sim $ a few units of Boltzmann's constant per baryon. If $\Delta s > 3$, nuclei in nuclear statistical equilibrium would be melted \cite{FMW} and at least some of the photo-dissociation burden on the shock could be alleviated. Whether by helping revive the stalled shock or by altering the thermal environment in its vicinity, the prospects for an explosion likely would be enhanced in this scenario, even in a simplistic one-dimensional supernova model.

The energy $E_s$ delivered by sterile neutrinos can vary greatly, depending on the core temperature, which need not be the same for all supernovae. Supernovae with $E_s \sim 10^{51} - 10^{52}$~erg should have a healthy shock wave and should be more likely to produce a neutron star, while supernovae with $E_s \ll 10^{51}$~erg are more likely to produce a black hole, due to a weaker shock and more infalling material. 
In any case, the existence of eosphoric sterile neutrinos with the properties we consider could alter the standard core collapse supernova energetics considerations while staying within existing observational bounds.

\section{Neutrino signature \label{nu signal}}

Eosphoric sterile neutrino decay will produce $\nu_\mu$ or $\nu_\tau$ neutrinos with energies $\epsilon_\nu \sim 80 \MeV$. Although very energetic, these neutrinos likely still will escape from the supernova. Their optical depth for scattering on the surrounding ambient matter is
\beq
\tau_\nu = G_F^2 \epsilon_\nu^2 \frac{\rho}{m_n} R < 10^{-2},
\eeq
for $\rho < 10^{5} {\rm g/cm^3}$ at $R \sim 10^9$~cm.  The optical depth could be somewhat greater for the case of mixing with tau neutrinos, because the decays of sterile neutrinos could take place closer to the core, as discussed above.

Neutrino observations from supernova explosions can help test the scenario presented in this paper.
In the event of a galactic supernova, experiments could detect the initial burst, 1-5 s duration, of
energetic $\nu_{\mu,\tau}$ from eosphoric neutrino decays, followed by a longer ($10\--15\,{\rm s}$)
neutrino signal of $\sim 15 \MeV$ neutrinos.  This signal will, of course, be modified by standard neutrino self-coupling
and matter-affected oscillations~\cite{Dighe:1999bi}. 
The detection probability of the very energetic
neutrinos is larger than that of ordinary neutrinos by a factor of $\sim (80 {\rm MeV}/15 {\rm
MeV})^2 = 28$. However, they carry only a small portion, $10^{51} {\rm erg} / 3\times 10^{53} {\rm
erg} \sim 0.3 \%$, of the total energy carried away by neutrinos. Thus, their time-integrated flux
will be a small fraction of the total neutrino flux: 
\beq (F_{80}/F_{\rm total}) \sim 28 \times 0.3 \% \sim 8\%  \label{detection} \eeq

The predicted neutrino signal is in agreement with the neutrino observations from SN1987A.
The Kamiokande detector observed 12 neutrino-induced events associated with SN1987A~\cite{SN1987A}. This implies that it could have seen one event originating from sterile neutrino decays.  This is not in contradition with the data~\cite{SN1987A}.  In addition, the emission of sterile 
neutrinos depends very sensitively on the core temperature,  growing as a high power of this quantity.  Since SN1987A  was an unusual  supernova, both from the standpoint of its progenitor star and the apparent absence of a pulsar in
the remnant,  it is also possible that the eosphoric neutrino production was inefficient due to a
lower than average temperature in the core. In this case an even lower flux of energetic
neutrinos would have been expected.  We emphasize, however, that even with the core temperature
expected in an average supernova, SN1987A should have produced less than one event stemming from
sterile neutrinos with the parameter ranges considered here.  The non-observation of gamma rays
from SN1987A~\cite{Kolb:1988pe} is consistent with our model because, in the core collapse of a blue supergiant, the gamma rays should have been absorbed by the envelope.

Modern and future detectors are expected to observe high neutrino-induced event rates from a galactic supernova.

\section{Conclusions}

Heavy sterile neutrinos could prove to be very important in compact astrophysical systems, such as supernovae, essentially because they could move significant amounts of energy around in ways that ordinary active neutrinos and hydrodynamic motions cannot.
In this paper, we have showed that sterile neutrinos with mass $\sim 200 \MeV$ and small mixing $\sin^2
\theta \sim 10^{-8}$ with either $\nu_\mu$ or $\nu_\tau$ could facilitate energy transport from the
supernova core to the shock front, possibly ultimately leading to a successful explosion, but in any case altering supernova energetics in ways which change the standard core collapse paradigm, yet produce signatures and behaviors that could remain within existing observational bounds.

The eosphoric sterile neutrino scenario can be tested by future neutrino observations from galactic
supernova explosions. A supernova should  produce a short burst of very energetic neutrinos followed
by a longer (and much more powerful) signal of lower-energy neutrinos. The non-observation of very
energetic neutrinos from SN1987A is consistent with our model because of the low flux. However, 
present and future neutrino detectors should be well positioned to detect the (less numerous)
80~MeV neutrinos. 
Observations of a Galactic supernova neutrino signal in these detectors could be used to constrain the eosphoric sterile neutrino parameter space in ways which could extend or be complimentary to future laboratory-based neutrino mixing probes.

\medskip

The work of A.K. and K.P. was supported in part by the DOE grant DE-FG03-91ER40662 and by the
NASA ATFP grant  NNX08AL48G; while G.M.F. was supported in part by NSF grant PHY-04-00359 at UCSD.

\end{document}